%% file: main.tex
\documentclass[conference]{IEEEtran}
\usepackage[english]{babel}
\usepackage{tikz}
\usepackage{amsmath,ctable,threeparttable,footnote,multirow,multicol}
\usepackage{longtable,color,psfrag,dsfont,supertabular,bm,units}
\usepackage{amssymb,amsmath}
\usepackage{graphicx} 
\usepackage{epstopdf}													
\usepackage{lastpage,algorithmic}
\usepackage{epsfig,amssymb}
\usepackage{bbding,cite}
\usepackage{multirow}
\usepackage{caption}
\usepackage{multicol}
\usepackage{subfigure}
\usepackage{url}
\usepackage[ruled,vlined,linesnumbered]{algorithm2e}
\usepackage{xcolor, colortbl}
\usepackage{xparse}
\usepackage{soul}
\usepackage{float}
\usepackage{comment}

\usetikzlibrary{automata,positioning,calc}

\usepackage{geometry}
\begin{document}

\IEEEoverridecommandlockouts
\IEEEpubid{\makebox[\columnwidth]{\copyright $\enskip$ 2018 IEEE \hfill }
\hspace{\columnsep}\makebox[\columnwidth]{\hfill }}

\title{Detection of Compromised Smart Grid Devices with Machine Learning and Convolution Techniques}

\author{\IEEEauthorblockN{Cengiz Kaygusuz, Leonardo Babun, Hidayet Aksu, and A. Selcuk Uluagac}
\IEEEauthorblockA{Cyber-Physical Systems Security Lab \\ Department of Electrical \& Computer Engineering,
Florida International University\\
10555 West Flagler St. Miami, FL 33174\\
Email:  \{ckayg001, lbabu002, haksu, suluagac\}@fiu.edu}
}

\maketitle
\IEEEpeerreviewmaketitle

\input{Abstract}
\input{Introduction}
\input{RelatedWork}
\input{AttackerModel}
\input{CLAnalysis}
\input{ProposedMethod}
\input{PrfEvl}
\input{Conclusions}
\input{Acknowledgement}
\bibliographystyle{IEEEtran}
\bibliography{bibtex}

\end{document}

%% file: Abstract.tex
\begin{abstract}

The smart grid concept has transformed the traditional power  grid into a massive cyber-physical system that depends on advanced two-way communication infrastructure to integrate a myriad of different smart devices. While the introduction of the cyber component has made the grid much more flexible and efficient with so many smart devices, it also broadened the attack surface of the power grid. Particularly, compromised devices pose great danger to the healthy operations of the smart-grid. For instance, the attackers can control the devices to change the behaviour of the grid and can impact the measurements. In this paper, to detect such misbehaving malicious smart grid devices, we propose a machine learning and convolution-based classification framework. Our framework specifically utilizes system and library call lists at the kernel level of the operating system on both resource-limited and resource-rich smart grid devices such as RTUs, PLCs, PMUs, and IEDs. Focusing on the types and other valuable features extracted from the system calls, the framework can successfully identify malicious smart-grid devices. In order to test the efficacy of the proposed framework, we built a representative testbed conforming to the IEC-61850 protocol suite and evaluated its performance with different system calls. The proposed framework in different evaluation scenarios yields very high accuracy (avg. 91\%) which reveals that the framework is effective to overcome compromised smart grid devices problem.

\end{abstract}

\begin{IEEEkeywords} 
Smart Grid, Compromised Devices, Cybersecurity, Machine Learning, Call List.
\end{IEEEkeywords}

%% file: Introduction.tex
\section{Introduction}

The ability to sense and react to what is happening in the power grid by smart devices 
has revolutionized the power industry. By measuring the grid parameters, smart grid devices are able to control the electrical grid much more safely and efficiently than ever before~\cite{6099519}. Indeed, the introduction of the smart devices into the decade-old power grid definitely is a giant step that
modernize the traditional grid; 
however, it also brings challenging security problems that are critical to tackle
~\cite{WANG20131344}. 

One of the most critical security problems in the power domain involves compromised smart grid devices. Compromising the smart devices such as sensors that measure the behavior of the power grid or controllers that either directly or indirectly controlling the behavior of the grid can have dire consequences: 
For instance, a sensor supplying false information may cause the control device to raise the voltage, possibly overloading the grid. Similarly,  a malicious activity on a control device may accomplish the same hazard directly, making electricity unavailable. To ensure a healthy supply of such a critical resource, it must be ensured the smart grid devices on the grid must behave as expected and provide healthy operations.

In this paper, we propose a new framework to detect compromised smart devices in a smart grid environment. Specifically, the framework extracts 
statistics of system and library calls at the kernel level in the operating system which is subsequently fed into a machine-learning based classification model and convolution process. Analyzing the detailed metrics of how two call lists differ on type, length, distribution, and ordering, the proposed framework is able to identify benign devices from the compromised ones in all the evaluated cases. 
In addition, the framework obtains high accuracy when applying it on the data gathered from a representative testbed of smart grid devices conforming to IEC61850 protocol suite. 

Our key contributions are listed as follows:
\begin{itemize}
    \item We propose a detection framework that combines information extracted from system and library call lists (type of calls, length of call lists, and ordering of the calls), convolution, and machine learning algorithms to identify compromised smart grid devices based on the devices' behavior.
    \item We propose an adversary model that considers three fundamental malicious activities stemming from compromised devices: direct grid control, indirect grid control, and surveillance activity from attackers.
    \item We demonstrate the efficacy of the proposed framework by evaluating 5 different realistic cases that specify how behaviour of authentic and compromised devices can differ in the smart grid. 
    \item Finally, we obtained high accuracy (91\% average) on the detection of compromised smart grid devices for all the different analyzed cases.
\end{itemize}

The remainder of the paper is structured as follows: Section~\ref{sec:related} presents the related work. Section~\ref{sec:smartgridcontext} discusses the smart grid context and explicitly defines the adversary model and the problem being  solved. Section~\ref{sec:clanalysis} describes the analysis of call list patterns. Section~\ref{sec:algo-overview} discusses how to make a distinction for each case. Section~\ref{sec:prf-evl} evaluates the performance of the framework and finally, Section~\ref{conclusion} concludes the paper and discusses future work.

%% file: RelatedWork.tex
\section{Related Work} \label{sec:related}

Existing works investigating the security of smart grid and industrial control systems (ICS) mostly deal with 
identifying counterfeit devices in the supply chain domain as the main focus for the compromised devices. In~\cite{counterfeitNetwork} and \cite{counterfeit6}, the authors offer different approaches for detecting fake electronic parts. Outbound beaconing, intelligent secure packaging, and better tracking systems are some of the countermeasures that are proposed to fight against counterfeiting on the supply chain side \cite{SEC_Project}. Common theme of these research is a focus on hardware level detection.  In a similar fashion, authors in~\cite{7996877} dealt with detecting counterfeit devices on software layer using statistical correlation. Simulating three different attack models on a smart grid testbed, the authors compared the data gathered from devices to ground truth. A hand-picked threshold value were used to decide whether the device was compromised.

On the other hand, the analysis of network traffic 
with the intention of classifying device behavior is a well studied subject.
Ahmed et al. presented a plethora of statistical and machine learning methods for behavioral network traffic analysis~\cite{AHMED201619}. Specifically in ICS, 
the authors of \cite{counterfeit2} are using network traffic to identify counterfeit ICS devices.  

The techniques that are employed in this paper are intersecting with some anomaly detection approaches. Existing data mining approaches has been reviewed by Agrawal et al.~\cite{AGRAWAL2015708}. In the smart grid context, authors in~\cite{anomalydetection} use a neural network model to detect malicious voltage control actions. In~\cite{attackdetection1}, researchers apply a rule based detection mechanism to detect attacks in smart grid. 

\vspace{3pt}\noindent\textbf{Difference from existing work-} \textit{Our work differs from the existing research as follows: The proposed framework operates on software level (at the kernel), and works for devices both in the supply chain as well as outside of it. Compared to approaches that solely focuses on the analysis of network data, the use of system and library call data offers a unique insight on the behavior of devices and is immune to random factors that affects the network state. To the best of our knowledge, this is the first study to provide a comprehensive framework with machine-learning and convolution techniques for the analysis of system and function call lists in the context of smart grid.}

%% file: AttackerModel.tex
\section{Smart Grid Context and Adversary Model} \label{sec:smartgridcontext}

This section discusses the adversary model focusing on the compromised device problem and the representation of the behavior of smart grid devices as a state machine. 

\subsection{Adversary Model}
As discussed earlier, the adversary model in this work deals with an unauthorized control of the smart grid devices via compromising.
In case of such a malicious event, the activity of a malicious actor could be categorized in three distinct classes: 

\begin{itemize}
    \item \textbf{Direct grid control with specific commands:} A compromised command \& control device, such as an IED, may allow the attacker to issue commands directly to affect the state of the grid.
    \item \textbf{Indirect grid control via fake measurements:} A compromised sensor may send fake measurements to indirectly exert control over the smart grid.
    \item \textbf{Surveillance of sensitive data:} A compromised device may allow sensitive and confidential data and measurements to be gathered from the devices.
\end{itemize}

\subsection{Characterization of Smart Grid Device Processes}
Majority of the critical field devices that are utilized in the grid include RTUs, PLCs, PMUs, and IEDs. In this work, the proposed framework specifically focus on these field devices:  

\begin{itemize}
    \item \textbf{Remote Terminal Unit (RTU):} Monitors relevant parameters about a system of subject and transmits this data to a central unit. This is an example of a \textit{resource limited} device, where memory and computation power is rather limited.
    \item \textbf{Programmable Logic Circuit (PLC):} Directly controls actuators to manipulate physical phenomena in the grid. 
    \item \textbf{Phasor Measurement Unit (PMU):} Measures electrical waves in an electrical grid.
    \item \textbf{Intelligent Electornic Devices (IED):} Receives data from sensors and issues control commands to regulate the grid. IED is an example of a \textit{resource-rich} device, where in contrast to resource-limited devices, memory and computation power is abundant.
\end{itemize}

\begin{figure}
\small
\centering
\begin{tikzpicture}[->, shorten >=5pt, shorten <= 5pt,node distance=3cm,on grid,auto] 
    \node[state]            (qI)                                         {Idle}; 
    \node[state]            (af)        [left =of qI]                    {Adjust Frequency};
    \node[state]            (av)        [right=of qI]                   {Adjust Voltage};

    \path   (qI.north)    edge [bend left]  node [right, yshift=0.4cm, xshift=-0.3cm]   {high voltage} (av.north)
            (av)    edge    node    {return}   (qI)
            (qI.north)    edge [bend right]  node [above, yshift=0.1cm] {frequency anomaly} (af.north)
            (af)    edge    node [below] {return} (qI);
\end{tikzpicture}
\caption{A partial state machine representing reactive nature of an IED. In case IED senses a high voltage, it does necessary computations and routine calls to adjust the voltage; and in case of a frequency anomaly, the device, then, adjusts frequency. This behavior could be generalized to other smart grid devices.} \label{figure-state-machine}
\vspace{-0.2in}
\end{figure}
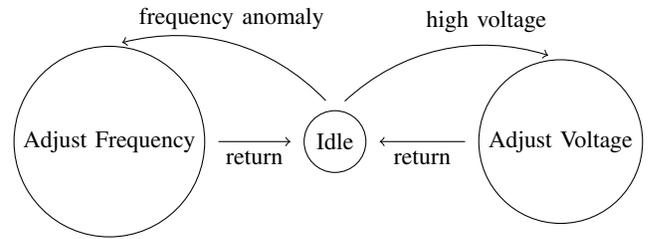

A key observation about these devices is that they are reactive: they respond to the events they receive in the smart grid in a deterministic fashion. Figure \ref{figure-state-machine} illustrates this behavior with a state machine representation. 
    
Moreover, most of the computation nowadays is done through programs that run on operating systems. A program must interact with the operating system to utilize system's resources through libraries that directly or indirectly use a standard library, such as allocating memory and sending a packet over the network to communicate with other computers. By obtaining system and/or library call traces over time, it is possible to identify which state the computing unit was operating on by analyzing the call list~\cite{1366573}. In other words, a computing unit responds to an event by a series of computation of which leaves a deterministic trace of system and library calls.

\subsection{Problem Definition}
After defining the adversary model and characterizing the smart grid devices in terms of state machines, here we introduce another state machine representation which serves as the explicit definition of the compromised behavior detection problem. 

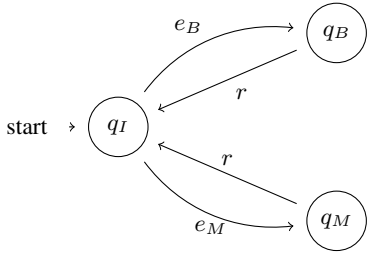
\begin{figure}
\small
    \centering
    \begin{tikzpicture}[->, shorten >=5pt, shorten <= 5pt, node distance=2.5cm, on grid, auto]
        \node[state] (qc) {$q_B$};
        \node[state] (qm) [below =of qc] {$q_M$};
        \node[state,initial] (qi) [left =of $(qc)!0.5!(qm)$] {$q_I$};

    \path   (qi) edge [bend left] node {$e_B$} (qc)
            (qc) edge node {$r$} (qi)
            (qi) edge [bend right] node[below] {$e_M$} (qm)
            (qm) edge node[above] {$r$} (qi);
    \end{tikzpicture}
    
    \caption{A state machine showing the initial, the benign computing, and the malicious (compromised) computation states. 
    }
    \vspace{-0.2in}
    \label{fig:st-mach-1}
\end{figure}

Let us consider the following state machine, which is a simplified version of the earlier state machine and shown in Figure~\ref{fig:st-mach-1}, where
\begin{itemize}
    \item $q_I$ is the idle state,
    \item $q_B$ is the \textit{benign computation mode}, e.g.,  expected activity. This is the state which the intended computation takes place; for example, a sensor might receive a timed IRQ to gather the readings and send it to a central unit,
    \item $q_M$ is the \textit{compromised mode}, e.g., unexpected activity. The device could be doing anything here, e.g. poisoning the sensor measurements or sending harmful control commands as discussed earlier.
\end{itemize}

The goal of this work is to identify whether the monitored device has ever operated in state $q_M$. With the assumption that every described state emits a deterministic trace of system and library call lists, it is possible to construct a framework that can identify whether a given device is compromised. 
Finally, it is assumed that benign and malicious activities do not produce identical call list patterns. 

%% file: CLAnalysis.tex
\section{Analysis of Call List Patterns} \label{sec:clanalysis}

To effectively utilize call lists as a discriminatory point between computing states, it is necessary to define what constitutes a call list and examine various measures of how two call lists are different. In this section, we define the call list and introduce the set of measures that are used in this work. Afterwards, the cases, which are based on the values of the defined metrics may take, are discussed.

\subsection{Definitions}

\textbf{A Call List $L_S$} is a finite sequential list of system or library calls when a computing unit finishes its operation in an arbitrary state $q_S$. 

\begin{equation}
\small
    SD\Bigg(
    \begin{vmatrix}
        malloc \\
        malloc \\
        free \\
        free
    \end{vmatrix},
    \begin{vmatrix}
        malloc\\
        free\\
    \end{vmatrix}
    \Bigg) = 0
    \label{eq:sd-ex}
\end{equation}

\textbf{Set Distance} $SD(L_1, L_2)$ is the measure of how two call lists are different according to the type of calls they inhibit. Let A be the set of calls in $L_1$, and B the set of calls in $L_2$. The function $SD(L_1, L_2)$ is simply the number of \textit{unique} elements (cardinality) that is contained in A, but not in B. Formally stated, $SD(L_1, L_2) = |A - B|$. Note that $SD(L_1, L_2) \neq SD(L_2, L_1)$. Equation~\ref{eq:sd-ex} gives an example where $SD = 0$.

\begin{equation}
\small
    LD\Bigg(
    \begin{vmatrix}
        malloc \\
        malloc \\
        malloc
    \end{vmatrix},
    \begin{vmatrix}
        free \\
        free \\
        free \\
    \end{vmatrix}
    \Bigg) = 0
    \label{eq:ld-ex}
\end{equation}

\textbf{Length Distance} $LD(L_1, L_2)$ is simply the difference of number of system calls contained by two call lists. $LD = 0$ indicates two call lists are of the same length, while $LD \neq 0 $ indicates one list is longer than another by given amount, without specifying which one it is. Equation~\ref{eq:ld-ex} gives an example where $LD = 0$. Note that $LD(L_1, L_2) = LD(L_2, L_1)$.

\begin{equation}
\small
    ED\Bigg(
    \begin{vmatrix}
        malloc \\
        malloc \\
        free \\
        free 
    \end{vmatrix},
    \begin{vmatrix}
        malloc\\
        free\\
        malloc \\
        free
    \end{vmatrix}
    \Bigg) = 0
    \label{eq:ed-ex}
\end{equation}

\textbf{Euclidean Distance} $ED(L_1, L_2)$ is a measurement unit that intermixes both type and length difference between two call lists. $v_{L_i}$ is an $N$ dimensional vector where each dimension is mapped to total number of calls made to that particular system or library function belonging to call list $L_i$. With this definition, $ED(L_1, L_2)$ is simply equal to $|v_{L_1} - v_{L_2}|$, or $ED(L_1, L_2) = |v_{L_1} - v_{L_2}|$. Equation~\ref{eq:ed-ex} gives an example where $ED = 0$.

\begin{equation}
\small
    HD\Bigg(
    \begin{vmatrix}
        malloc \\
        malloc \\
        free \\
        free 
    \end{vmatrix},
    \begin{vmatrix}
        malloc\\
        free\\
        malloc \\
        free
    \end{vmatrix}
    \Bigg) = 2
    \label{eq:hd-ex}
\end{equation}

\textbf{Hamming Distance} $HD(L_1, L_2)$ is simply the number of operations required to be undertaken in order to make two call lists identical. Note that $HD(L_1, L_2) = 0$ implies two lists are identical. Equation~\ref{eq:hd-ex} gives an example where $HD = 2$.

\subsection{Call List Difference Cases} \label{cases}
The following is the list of all the identified critical points. These are also utilized in the performance evaluations in Section~\ref{sec:prf-evl}. 

\begin{enumerate}
    \item $SD(L_M, L_C) > 0$: Malicious state makes a call that is not contained in the benign state.
    \item $SD(L_C, L_M) < 0$: Benign state makes a call that is not contained in the malicious state. Subsequent cases assume $SD(L_C, L_M) = SD(L_M, L_C) = 0$
    \item $LD(L_C, L_M) \neq 0$: Two call lists inhibit same type of calls, but differing in length. Subsequent cases assume $LD(L_C, L_M) = 0$
    \item $ED(L_C, L_M) \neq 0$: Two call lists are of the same length and contains the same type of calls, but their internal distribution is different. Next case assumes $ED(L_C, L_M) = 0$
    \item $HD(L_C, L_M) \neq 0$ Two call lists are of the same length, contains same type of calls, their internal distribution is the same but their order is not identical.
\end{enumerate}

%% file: ProposedMethod.tex
\section{Overview of the Framework} \label{sec:algo-overview}

\begin{figure}[t]
    \centering
    \includegraphics[scale=0.5]{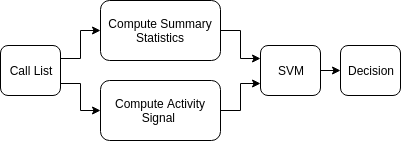}
    \caption{Overview of the framework. Using the harvested system or library call list, summary statistics and activity signal values are computed, which subsequently fed into a Support Vector Machine (SVM) to make a decision.}
    \vspace{-0.2in}
    \label{fig:overview}
\end{figure}

In this section, details of the proposed framework is introduced. As shown in Figure~\ref{fig:overview}, 
the framework utilizes a classifier model to make a decision whether or not the observed device has been exhibiting malicious activity (i.e., compromised device behaviour). The feature set contains the following classes:

\begin{enumerate}
    \item Total number of calls made for each call type,
    \item Average number of calls for each call type,
    \item A value derived from activity signal.
\end{enumerate}

As the collected behavioral data exhibits linearly separable features, a binary Support Vector Machine (SVM) classifier with a linear kernel is used.

\begin{figure*}
\small
    \centering
    \begin{tikzpicture}
        \tikzstyle{process} = [rectangle, minimum width=3cm, minimum height=1cm, text centered, draw=black]
        \tikzstyle{empty} = [rectangle, minimum width=3cm, minimum height=1cm, text centered, draw=black]

        \node(start) [text width=5, minimum width=2cm] {
            malloc\\
            malloc\\
            free\\
            free\\
            recv\\
            ...
        };
        \node(second) [right=of start] {
         \begin{tabular}{||c c c c||} 
             \hline
             \textbf{1} & \textbf{2} & \textbf{3} & \textbf{target} \\ [0.5ex] 
             \hline\hline
             malloc & malloc & free & free \\ 
             \hline
             malloc & free & free & usleep \\
             \hline
             ... & ... & ... & ... \\
             \hline
            \end{tabular}
        };
        \node(third) [right=of second] {[0, 1, 0, 1, ...]};
        \node(fourth) [right=of third] {[2, 5, 67, 68, ...]};
        \node(fifth) [right=of fourth] {68};
        
        \begin{scope}[->, shorten <= 9pt, shorten >= 9pt]
            \draw (start.north) to [bend left] node[above] {Bucketing} (second.north) ;
            \draw (second.north) to [bend left] node[above, yshift=5pt] {Prediction} (third.north);
            \draw (third.north) to [bend left] node[above] {Convolution} (fourth.north);
            \draw (fourth.north) to [bend left] node[above] {Max-Pooling} (fifth.north);
        \end{scope}

    \end{tikzpicture}
    \caption{Demonstration of how data is transformed when computing the activity signal value. The received call list is first pre-processed through a procedure called \textit{bucketing}. The pre-processed data is then fed into the classification model, which emits a 0 for correct and 1 for incorrect prediction. Resulting array of binary values is then sum-convoluted with a kernel size of 100. The convoluted values are then max-pooled with a non-overlapping sliding window of size 100. The output of the max-pooling is again processed with another round of convolution and max-pooling, which is continued until a single integer is obtained.}
    \vspace{-0.2in}
    \label{fig:activity-signal-flow}
\end{figure*}
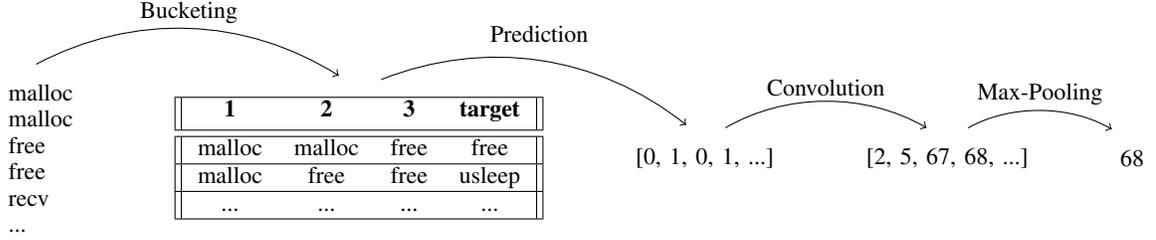

\textit{Activity signal} is the measure of how the order of calls in the observed list is in compliance with ground truth. Two call lists cannot be separated using summary statistics if two call lists differ only in the order of calls made. To address this challenge, Algorithms 1, 2, and 3 are proposed to be used in succession. Figure \ref{fig:activity-signal-flow} gives an overview on the data flow in the context of activity signal.
\begin{algorithm}[t]
\small
    \SetKwData{Left}{left}\SetKwData{This}{this}\SetKwData{Up}{up}
    \SetKwFunction{Union}{Union}\SetKwFunction{FindCompress}{FindCompress}
    \SetKwInOut{Input}{input}\SetKwInOut{Output}{output}
    \SetKw{KwBy}{by}
    \DontPrintSemicolon
    \caption{Bucketing}
    \label{algo:bucket}

    \Input{system or library call list}
    \Output{bucketed call list}
    
    \Begin{
        {$S_{bucket}$ $\leftarrow$ configure()}\;
        {$S_{window}$ $\leftarrow$ configure()}\;
        {T $\leftarrow$ configure()}\;
        
        \For{$i \leftarrow 1$ \KwTo input.length - $S_{bucket}$}{
            \For{j $\leftarrow$ 1 to $S_{bucket}$ - 1}{
                R[i][j] {$\leftarrow$} input[i + j - 1]\;
            }
            R[i].last {$\leftarrow$} input[i + $S_{bucket}$ - 1] \;
        }
        \KwRet{R}\;
    }
\end{algorithm}

The main idea is to use a classification model that predicts the next sequence of calls by inspecting the previous list of calls. If the predictions are true, it is inferred the call list is exhibiting expected patterns. On the other hand, if the predictions are failing, it is inferred the call list is exhibiting unexpected patterns; thus, the framework concludes the malicious activity is present. To quantify the mis-prediction ratio, a cascade of convolution and max-pooling operations are employed in the framework. 

\begin{algorithm}[t]
\small
    \SetKwData{Left}{left}\SetKwData{This}{this}\SetKwData{Up}{up}
    \SetKwFunction{Union}{Union}\SetKwFunction{FindCompress}{FindCompress}
    \SetKwInOut{Input}{input}\SetKwInOut{Output}{output}
    \SetKw{KwBy}{by}
    \DontPrintSemicolon
    \caption{Constructing Raw Prediction Signal}
    \label{algo:pred-signal}

    \Input{bucketed call list}
    \Output{raw prediction signal}
    
    \Begin{
        {$S_{bucket}$ $\leftarrow$ configure()}\;
        {$S_{window}$ $\leftarrow$ configure()}\;
        {T $\leftarrow$ configure()}\;
        
        \For{$i \leftarrow 1$ \KwTo R.length}{
            $C_{target}$ {$\leftarrow$} R[i].last \;
            $C_{predict}$ {$\leftarrow$} predict(R[i]) \;
            \eIf{ $C_{predict}$ = $C_{target}$}{
                P[i] {$\leftarrow$} 0 \;
            }{
                P[i] {$\leftarrow$} 1 \;
            }
        }
        
        \KwRet{P}\;
    }
\end{algorithm}

Upon receiving the call list, the first step is to pre-process the call list through an operation called \textit{bucketing}. Algorithm~\ref{algo:bucket} describes the computation required for this operation. Over a sliding window of a configured size called \textit{bucket length}, each call is mapped into one feature column, with the last one being the call the framework is trying to predict. The experimental results indicated a bucket length of 32 calls is an acceptable trade-off between run-time performance and accuracy.  

\begin{algorithm}[t]
\small
    \SetKwData{Left}{left}\SetKwData{This}{this}\SetKwData{Up}{up}
    \SetKwFunction{Union}{Union}\SetKwFunction{FindCompress}{FindCompress}
    \SetKwInOut{Input}{input}\SetKwInOut{Output}{output}
    \SetKw{KwBy}{by}
    \DontPrintSemicolon
    \caption{Reducing Raw Prediction Signal}
    \label{algo:reduction}

    \Input{raw prediction signal}
    \Output{an integer giving a measure of how far the call list is straying away from ground truth.}
    
    \Begin{

        {$T_C$} {$\leftarrow$} input\;
        \While{$T_C.length < S_{window}$}{
            \For{$i \leftarrow 1$ \KwTo $T_C.length - S_{window}$}{
                $T_A[i]$ $\leftarrow$ sum from $T_C[i]$ to $T_C[i+S_{window}]$\;
            } 
            
            \For{$i \leftarrow$ 1 \KwTo $T_A.length - S_{window}$ \KwBy $S_{window}$}{
                $T_B[i]$ $\leftarrow$ max from $T_A[i]$ to $T_A[i+S_{window}]$\;
            }
            $T_C$ $\leftarrow$ $T_B$\;
        }
        \KwRet $sum(T_C)$
    }

\end{algorithm}

After the bucketing procedure, the pre-processed data is fed into the prediction signal generator. Algorithm~\ref{algo:pred-signal} describes the computation required to accomplish this operation. For each entry resulting in the dataset after bucketing operation, the predictor yields 0 if it was correctly predicted, 1 if the prediction failed. The resulting array of binary values is named as \textit{activity signal}. For this process, the framework is utilizing random forest machine learning algorithm as the predictor as it provides the best performance. 

Lastly, Algorithm~\ref{algo:reduction} reduces the binary array obtained from Algorithm~\ref{algo:pred-signal} to a single integer value using a cascade of convolution and max-pooling. The convolution kernel is simply sums all the values in the sliding window. Max-pooling picks the maximum element over a sliding window. Since max-pooling procedure uses non-overlapping windows, the data size is reduced by a factor of \textit{window size} each time this operation is conducted. The resulting value is aimed to be higher in unexpected activity, and lower in expected activity. The cascade of convolution and max-pooling idea originates from convolutional neural networks (CNN), where a much more complex architecture involving these techniques are used for image recognition tasks with success~\cite{NIPS2012_4824}. 


%% file: PrfEvl.tex
\section{Performance Evaluation} \label{sec:prf-evl}

\begin{figure}
 \vspace{-0.15in}
    \centering
    \includegraphics[scale=0.5]{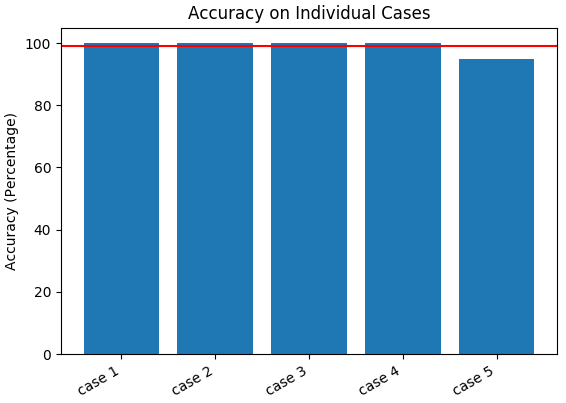}
    \caption{Accuracy of the framework applied individually to each case as defined in Section~\ref{cases}. The framework was able to exploit call list differences in all the cases. The red line indicates the average accuracy of 99\%.}
    \vspace{-0.30in}
    \label{fig:theo-acc}
\end{figure}

\begin{figure}
 \vspace{-0.10in}
    \centering
    \includegraphics[scale=0.5]{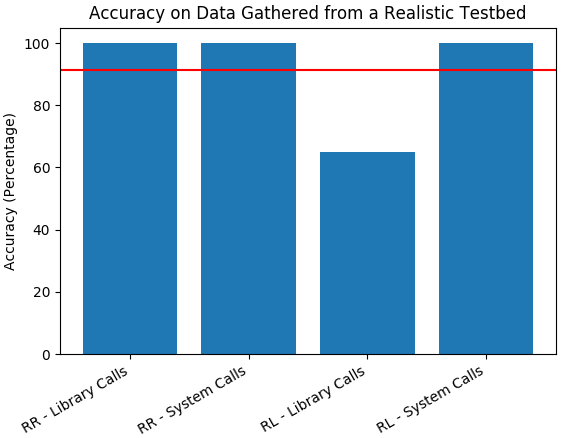}
    \caption{Results on data gathered from 
    resource rich (RR) devices and resource limited (RL) devices. The framework 
    accurately identifies compromised devices when looking solely at library calls, which indicate one of the two data sources may contain discriminatory information while the other one does not. The red line indicates the average accuracy of 91.25\%}
    \vspace{-0.25in}
    \label{fig:end-results}
\end{figure}

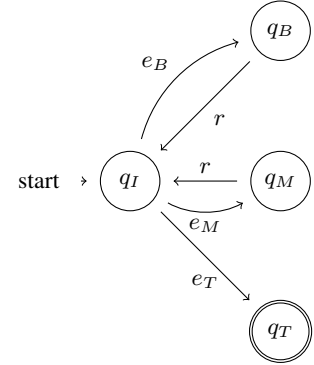
\begin{figure}
    \centering
\small
    \begin{tikzpicture}[->, shorten >=5pt, shorten <= 5pt, node distance=2cm, on grid, auto]
        \node[state, initial] (qi) {$q_I$};
        \node[state] (qm) [right =of qi] {$q_M$};
        \node[state] (qb) [above =of qm] {$q_B$};
        \node[state,accepting] (qt) [below =of qm] {$q_T$};
        
    \path   (qi) edge [bend left] node {$e_B$} (qb)
            (qb) edge node {$r$} (qi)
            (qi) edge [bend right] node[below] {$e_M$} (qm)
            (qm) edge node[above] {$r$} (qi)
            (qi) edge node[below, yshift=-4pt] {$e_T$} (qt);
    \end{tikzpicture}

    \caption{State machine representation of the test cases on a smart grid device.
    }
    \vspace{-0.2in}
    \label{fig:test-state-machine}
\end{figure}

To test the overall classifier against the cases identified in Section~\ref{sec:clanalysis}, the state machine depicted in Figure~\ref{fig:test-state-machine} was utilized to generate data representing the behavior of authentic and compromised devices. Each execution of the state machine on the smart grid device is called an \textit{experiment}. This state machine operates on \textit{events}. Given an event, the smart grid device makes a transition to the state concerned, emits a call list, and returns to the idle state.

The smart grid device's state machine operated on the following principles:

\begin{itemize}
    \item The machine starts at $q_I$.
    \item At each turn, the machine transitions to either state $q_C$ or $q_M$, with state transition probabilities $p_C$ and $p_M$. 
    \item At state $q_C$ and $q_M$, the machine outputs a list of pre-determined call list and transitions back into $q_I$.
    \item The total amount of events to be run in an experiment is randomly chosen according to a Gaussian distribution of mean of 10000 and standard deviation of 3000 events. A probability distribution is used to introduce variance in total number of calls. Mean and standard deviation numbers are picked by trial and error to ensure amount of data doesn't affect the outcome of the experiments. 
    \item After the total amount of events are processed, the machine transitions to terminal state $q_T$ and halts without producing any call trace.
\end{itemize}

For each case to be mentioned, there are a total of 60 experiments: 
\begin{itemize}
    \item 30 experiments run with $p_C = 1$ and $p_M = 0$, and the resulting data are assumed to be coming from authentic devices.
    \item 30 experiments run with $p_C = 0.99$ and $p_M = 0.01$. 
\end{itemize}

The machine learning classification model (i.e., random forest) that is dealing with the generation of activity signal is trained on one single instance of the authentic call list. Then, the decision model is trained with $2/3$ of the experiments, and the remaining $1/3$ is used for testing. 

The results of aforementioned setting is presented in Figure~\ref{fig:theo-acc}. 
In cases 1 through 5, the framework was able to exploit differences in call lists to accurately identify authentic devices from compromised ones. In case 6, the framework did as good as randomly guessing the device authenticity as benign and malicious computation emit completely identical call lists.

The framework is further tested on the dataset obtained in our previous study \cite{7996877}. The data was obtained using a representative smart grid implemented by utilizing open source IEC61850 library \textit{libiec61850}. The dataset consists of data obtained from resource-rich devices as well as resource-limited devices after emulating three explicit attacks: information leakage, measurement poisoning, and saving data into the device memory to be sent later to attackers. Devices are also run without emulating any malicious activity, resulting in call lists data that is utilized as ground truth. As with previous test, $2/3$ of the dataset is utilized for training, and remaining $1/3$ is used for testing.

The results are shown in Figure~\ref{fig:end-results}. The algorithm was able to make good use of both system and library call lists in resource rich devices. Though, it was only able to obtain a high accuracy using only system calls in resource-limited devices. This fact indicates one of the two data sources may contain discriminatory information while the other one does not. 

%% file: Conclusions.tex
\section{Conclusions and Future Work} \label{conclusion}
Compromised devices pose great danger to the healthy operations of the smart grid. The attackers may utilize compromised devices to change the behaviour of the grid and modify critical measurements. In this paper, we proposed a novel detection framework that combines information extracted from system and library call lists, convolution, and machine learning algorithms to detect compromised smart grid devices. The performance of the proposed framework on a realistic smart-grid testbed conforming to the IEC-61850 protocol suite 
was evaluated on 5 different realistic cases. The test cases specified how behaviour of authentic and compromised devices could differ in the smart grid. The evaluation results demonstrated that the proposed framework can perform with very high accuracy (average 91\%) on the detection of compromised smart grid devices. Although the proposed framework yielded highly accurate results, we ill consider other features from the devices to enhance to framework in our future work.  

%% file: Acknowledgement.tex
\section{Acknowledgements}

This material is based upon work supported by the Department of Energy under Award Number DE-OE0000779.

%% file: main.bbl
\begin{thebibliography}{10}
\providecommand{\url}[1]{#1}
\csname url@samestyle\endcsname
\providecommand{\newblock}{\relax}
\providecommand{\bibinfo}[2]{#2}
\providecommand{\BIBentrySTDinterwordspacing}{\spaceskip=0pt\relax}
\providecommand{\BIBentryALTinterwordstretchfactor}{4}
\providecommand{\BIBentryALTinterwordspacing}{\spaceskip=\fontdimen2\font plus
\BIBentryALTinterwordstretchfactor\fontdimen3\font minus
  \fontdimen4\font\relax}
\providecommand{\BIBforeignlanguage}[2]{{%
\expandafter\ifx\csname l@#1\endcsname\relax
\typeout{** WARNING: IEEEtran.bst: No hyphenation pattern has been}%
\typeout{** loaded for the language `#1'. Using the pattern for}%
\typeout{** the default language instead.}%
\else
\language=\csname l@#1\endcsname
\fi
#2}}
\providecommand{\BIBdecl}{\relax}
\BIBdecl

\bibitem{6099519}
X.~Fang, S.~Misra, G.~Xue, and D.~Yang, ``Smart grid -- the new and improved
  power grid: A survey,'' \emph{IEEE Communications Surveys Tutorials},
  vol.~14, no.~4, pp. 944--980, Fourth 2012.

\bibitem{WANG20131344}
\BIBentryALTinterwordspacing
W.~Wang and Z.~Lu, ``Cyber security in the smart grid: Survey and challenges,''
  \emph{Computer Networks}, vol.~57, no.~5, pp. 1344 -- 1371, 2013. [Online].
  Available:
  \url{http://www.sciencedirect.com/science/article/pii/S1389128613000042}
\BIBentrySTDinterwordspacing

\bibitem{counterfeitNetwork}
S.~Sathyanarayana, W.~H. Robinson, and R.~Beyah, ``A network-based approach to
  counterfeit detection.'' in \emph{IEEE International Conference on
  Technologies for Homeland Security}, ser. HST, Waltham, Massachusetts, 2013,
  NS.

\bibitem{counterfeit6}
{A. Kanovsky, P. Spanik and M. Frivaldsky}, ``{Detection of electronic
  counterfeit components},'' in \emph{2015 16th Int. Scientific Conf. on
  Electric Power Engineering (EPE)}.\hskip 1em plus 0.5em minus 0.4em\relax
  Kouty nad Desnou: IEEE, May 2015, pp. 701 -- 705.

\bibitem{SEC_Project}
\BIBentryALTinterwordspacing
{D. van Opstal, U.S. Resilience Project}, ``{Supply chain solutions for smart
  grid security: Building on business best practices.}'' Sep 2012. [Online].
  Available:
  \url{http://usresilienceproject.org/wp-content/uploads/2014/09/report-Supply_Chain_Solutions_for_Smart_Grid_Security.pdf}
\BIBentrySTDinterwordspacing

\bibitem{7996877}
L.~Babun, H.~Aksu, and A.~S. Uluagac, ``Identifying counterfeit smart grid
  devices: A lightweight system level framework,'' in \emph{2017 IEEE
  International Conference on Communications (ICC)}, May 2017, pp. 1--6.

\bibitem{AHMED201619}
M.~Ahmed, A.~N. Mahmood, and J.~Hu, ``A survey of network anomaly detection
  techniques,'' \emph{Journal of Network and Computer Applications}, vol.~60,
  no. Supplement C, pp. 19 -- 31, 2016.

\bibitem{counterfeit2}
U.~Guin, D.~Forte, and M.~Tehranipoor, ``Anti-counterfeit techniques: From
  design to resign,'' in \emph{Proceedings of the 2013 14th International
  Workshop on Microprocessor Test and Verification}.\hskip 1em plus 0.5em minus
  0.4em\relax Washington, DC, USA: IEEE Computer Society, 2013, pp. 89--94.

\bibitem{AGRAWAL2015708}
\BIBentryALTinterwordspacing
S.~Agrawal and J.~Agrawal, ``Survey on anomaly detection using data mining
  techniques,'' \emph{Procedia Computer Science}, vol.~60, no. Supplement C,
  pp. 708 -- 713, 2015, knowledge-Based and Intelligent Information \&
  Engineering Systems 19th Annual Conference, KES-2015, Singapore, September
  2015 Proceedings. [Online]. Available:
  \url{http://www.sciencedirect.com/science/article/pii/S1877050915023479}
\BIBentrySTDinterwordspacing

\bibitem{anomalydetection}
A.~M. Kosek, ``Contextual anomaly detection for cyber-physical security in
  smart grids based on an artificial neural network model,'' in \emph{2016
  Joint Workshop on Cyber- Physical Security and Resilience in Smart Grids
  (CPSR-SG)}, April 2016, pp. 1--6.

\bibitem{attackdetection1}
Y.~Sun, X.~Guan, T.~Liu, and Y.~Liu, ``A cyber-physical monitoring system for
  attack detection in smart grid,'' in \emph{2013 IEEE Conference on Computer
  Communications Workshops (INFOCOM WKSHPS)}, April 2013, pp. 33--34.

\bibitem{1366573}
M.~M. Yasin and A.~A. Awan, ``A study of host-based ids using system calls,''
  in \emph{2004 International Networking and Communication Conference}, June
  2004, pp. 36--41.

\bibitem{NIPS2012_4824}
A.~Krizhevsky, I.~Sutskever, and G.~E. Hinton, ``Imagenet classification with
  deep convolutional neural networks,'' in \emph{Advances in Neural Information
  Processing Systems 25}, F.~Pereira, C.~J.~C. Burges, L.~Bottou, and K.~Q.
  Weinberger, Eds.\hskip 1em plus 0.5em minus 0.4em\relax Curran Associates,
  Inc., 2012, pp. 1097--1105.

\end{thebibliography}
